\documentclass[10pt,letterpaper]{article}
\usepackage[top=0.85in,left=2.75in,footskip=0.75in]{geometry}

\usepackage{amsmath,amssymb}

\usepackage{changepage}

\usepackage[utf8x]{inputenc}

\usepackage{textcomp,marvosym}

\usepackage{cite}

\usepackage{nameref,hyperref}

\usepackage[right]{lineno}

\usepackage{microtype}
\DisableLigatures[f]{encoding = *, family = * }

\usepackage[table]{xcolor}

\usepackage{array}

\newcolumntype{+}{!{\vrule width 2pt}}

\newlength\savedwidth

\raggedright
\usepackage[aboveskip=1pt,labelfont=bf,labelsep=period,justification=raggedright,singlelinecheck=off]{caption}

\makeatletter
\renewcommand{\@biblabel}[1]{\quad#1.}
\makeatother

\usepackage{lastpage,fancyhdr,graphicx}
\usepackage{epstopdf}
\fancyheadoffset[L]{2.25in}
\fancyfootoffset[L]{2.25in}
\begin{document}
\vspace*{0.2in}

\begin{flushleft}
{\Large
\textbf\newline{Changes in mobility patterns in Europe during the COVID-19 pandemic: Novel insights using open source data} 
}
\newline
\\
Anna Sigrídur Islind\textsuperscript{1\Yinyang}
María Óskarsdóttir\textsuperscript{1\Yinyang*}
Harpa Steingrímsdóttir\textsuperscript{1\Yinyang}
\\
\bigskip
\textbf{1} Department of Computer Science, Reykjavík University, Reykjavík, Iceland
\\
\bigskip

%
%
\Yinyang These authors contributed equally to this work.





* mariaoskars@ru.is

\end{flushleft}
\section*{Abstract}
The COVID-19 pandemic  has changed the way we act, interact and move around in the world. The pandemic triggered a worldwide health crisis that has been tackled using a variety of strategies across Europe. Whereas some countries have taken strict measures, others have avoided lock-downs altogether. In this paper, we report on findings obtained by combining data from different publicly available sources in order to shed light on the changes in mobility patterns in Europe during the pandemic. Using that data, we show that mobility patterns have changed in different counties depending on the strategies they adopted during the pandemic. Our data shows that the majority of European citizens walked less during the lock-downs, and that, even though flights were less frequent, driving increased drastically. In this paper, we focus on data for a number of countries, for which we have also developed a dashboard that can be used by other researchers for further analyses. Our work shows the importance of granularity in open source data and how such data can be used  to shed light on the effects of the pandemic.


\section*{Introduction}
In December 2019, a severe respiratory disease emerged in Whuan, China \cite{wu2020new}. Since the first identified case of the virus COVID-19 was reported on 8. December 2019 \cite{huang2020clinical,zhu2020novel}, the virus has spread quickly, causing a worldwide health crisis. Since the crisis broke, different types of crisis management strategies have been actualised, impacting the way people in different countries act and interact in ways that have not yet fully been realised \cite{shangguan2020caused}. The pandemic has affected people living in various countries in different ways, since each country has had its own approach and policy for handling the crisis situation. This goes for testing strategies, requirements around quarantine, imposing lock-downs, social-distancing and closing of schools to name a few \cite{kissler2020social}. Additionally, as the virus was spreading around the globe, countries were effected by it to a different extent, and took longer or shorter time to respond to the imminent threat.
In Europe alone, the response to the pandemic has varied greatly. For example, although Spain and the United Kingdom both reported their first COVID-19 case on January 31, Spain imposed a lock-down nine days before the United Kingdom (43 and 52 days, respectively, after the first diagnosed case). In contrast, Norway and Denmark reported their first cases on February 26 and 27, and waited only 15 days to significantly restrict the travels to, from and within their countries by imposing a lock-down. Since these measures were taken, it has become clear that non-pharmaceutical interventions significantly and substantially slowed the growth of the pandemic \cite{hsiang2020effect}.

Previous studies have shown the benefits of developing and researching information systems with the aim of pursuing grand challenges such as natural disasters, public health issues, dealing with poverty, climate change and the ageing society \cite{braa2004networks,aagerfalk2020information,heeks2010social,rajao2009conceptions,tim2017digitally,islind2018co,pan2020fighting}. 
Following that stream of research on grand challenges, and in light of the new grand challenge that we are now facing with the ongoing pandemic, we suggest that providing knowledge and developing information systems based on open source data can provide new insights on the pandemic and its effects. There is a related call for research on the aggregated flow of people during the pandemic in order to show the impact of different measures taken during the state of crisis \cite{buckee2020aggregated}. This paper directly contributes to that call, and that gap in the literature. 

In this paper we more specifically report on our findings from using open source data from different resources, including the Apple Maps Mobility Trend data set and the COVID-19 Flight data set provided by OpenSky, for the purpose of developing a dashboard which shows the changes in mobility patterns in Europe and analyse the data with the purpose of shedding light on mobility trends following the pandemic. Our research questions are two-fold: i) How has the COVID-19 pandemic changed mobility patterns in Europe; and ii) What conclusions can be drawn from open source data about the impact of the COVID-19 pandemic on well-being of individuals in Europe in relation to strategy steps taken by the various countries. Using the data and our dashboard, we show the impact of the different strategies put into place in Europe during the pandemic and show how mobility patterns changed in different counties. The findings show how the majority of Europeans walked less during the lock-downs and social distancing periods, and that even though flights were less frequent, driving increased drastically instead. The main contribution of our research is firstly an analysis of mobility patterns in relation to measures taken to mitigate the effects of the pandemic in different countries in Europe and secondly an information system --the dashboard-- with data and visualisations showing changes in mobility patterns in Europe. The dashboard will be made available online upon publication of this paper. Additionally, we illustrate how novel use of open source data can elucidate various aspects of the pandemic which indicates the importance of granularity in open source data, and the significance of updating open source data frequently so that it can be used to tackle grand challenges such as this one.

The rest of this paper is organized in the following way. In the next section we discuss the related literature on mobility, crisis management, predicting the spread of the pandemic and the effect of imposed restriction on general population, followed by an explanation of our proposed methods and materials in Section \ref{s:mm}.  In Section \ref{s:result} we present the results of our analyses and discuss their implications in Section \ref{s:discussion}. We conclude the paper with some final remarks and directions for future research in Section \ref{s:conclusion}.

\section*{Related Work}
Human mobility has been a popular research topic for over a decade. With increased availability of granular, geo-located mobility data research has revealed that people tend to follow simple and reproducible patterns; they are likely to return to a few highly frequented locations and their trajectories are highly regular both in terms of time and space \cite{gonzalez2008understanding}. Recently, there is a growing interest in understanding mobility patterns and social interactions in healthcare situations through a data-driven approach \cite{oskarsdottir2020interaction}. Such data has also been used to assess population displacement following disasters \cite{wilson2016rapid}. In light of that and with the COVID-19 disease spreading around the globe, mobility research has gain new heights, whereby researchers, on the one hand, strive to understand and predict the spread of the virus using geo-located data, and, on the other hand, to understand the effects of the pandemic on human mobility through the use of data. 

The first line of research is focused on understanding and predicting how the virus spreads, which includes estimating its basic reproduction number $R_0$. Over the years, advanced epidemic and compartment models that rely on detailed mobility data have been developed to simulate and investigate epidemic spread. These and alternative techniques have successfully been used for H1N1 as well as seasonal influenza. \cite{balcan2010modeling,bajardi2011human,kishore2020flying}. With a new, unknown infectious disease, estimation of the basic reproduction number often outlines a first step, and furthermore to model the virus's spread and effect of management strategies.
Early studies of the spread of COVID-19 in China and its origin in Wuhan, indicated that the basic reproduction number could be as high as $R_0=6.47$, and that with strict travel restrictions, the disease would peak in about 2 weeks time \cite{tang2020estimation}. In an effort to estimate the latent infection ratio in Wuhan before the lock-down, a study showed that a significant number of infected people had left the city at that time, and gave an estimate of $R_0=3.24 $\cite{cao2020incorporating}. Strict control measures proved effective in China. The lock-down in Wuhan delayed the spread of the virus to other cities in China by almost 3 days, and cities that implemented restrictions preemptively reported fewer cases in the first weeks than cities that started control later \cite{tian2020investigation}. 
An advanced epidemic simulation model with real-time mobility data from Wuhan furthermore established that the drastic control measures substantially mitigated the spread of the virus \cite{kraemer2020effect}.
Moreover, taking global mobility data into account, including flights,  showed that the travel ban in Wuhan only postponed the spread by 3-5 days in China, but case importation of infections globally was reduced by 80\% until mid February \cite{chinazzi2020effect}. 
Researchers are currently also starting to use geo-located mobility data coupled with census and demographic data in the USA to predict the effects of a second wave of the pandemic, concluding that social distancing, high levels of testing, contact-tracing and household quarantine will be most effective in keeping the infection rate low \cite{aleta2020modelling}.

The second viewpoint of using mobility data to understand the effects of the pandemic on human mobility, has been researched to a less extent. However, even though the current literature on the effects of COVID-19 on mobility patterns shows the detection of dramatic changes in mobility in the wake of COVID-19, most of the research is focused on the changes within the USA \cite{warren2020mobility}, in Italy \cite{carteni2020mobility}, and in China \cite{jia2020population}. Neither analysing more general mobility trends cross continents nor a general level comparison between countries have thereby been attempted to date. 

Bushfires, tornadoes, avalanches, floods and earthquakes are all non-routine events that can be classified as unpredictable extreme situations which call for shifts from routine practice \cite{schmidt2013behaviour}. These types of situations, put unfamiliar demands and strains on both individuals, and on policy-makers \cite{karanasios2020gatekeepers,wilson2016rapid}. What has characterised the situation surrounding COVID-19 is that no-one was prepared \cite{spinelli2020covid}. As a result, we can see the situation surrounding the COVID-19 pandemic as a crisis situation due to its unpredictable non-routine nature where different countries have activated different strategies for handling the crisis. 
The crisis handling of COVID-19 has an impact on mobility patterns since a lock-down is bound to affect how people move about in that particular country in their day to day lives. The different countries have enforced and followed different rules and policies. When comparing the different strategies, the research to date has for instance showed the effects of the different reactions, and crisis management strategies. For instance, \cite{flaxman2020estimating} investigate the effect of major inventions in eleven European countries from February to May. They show that interventions made at the time were sufficient to drive basic reproduction number ($R_0$) below one and thus gain control of the pandemic.  According to their study, between 3.2\% to 4\% of the inhabitants in those eleven countries had been infected by May 4. In their study, they show that lock-downs in particular had a large impact on reducing transmission.

The literature on the basic reproduction number and on which strategies have proven most effective is growing and will continue to grow as more and more is known about the virus, its spread as well as short and long term effects. Only when the whole pandemic period is examined will the effects of the pandemic be clear. However, the literature listed herein gives a view of the first wave of the strategies, spread and mobility during the first responds of crisis mode. 

\section*{Methods \& Materials}\label{s:mm}

In our study, we use only open source and publicly available data. As the pandemic was spreading and to this day, several organisations are sharing aggregated data publicly, often with daily updates. Furthermore, individual counties share additional and specific data and information for their citizens.

To investigate changes in mobility, we considered walking, driving and flying as means of travel.  For the first two, we use Apple Map Mobility Trends (MT) \footnote{See: \href{https://www.apple.com/covid19/mobility}{https://www.apple.com/covid19/mobility}} which show the changes in routing requests from day to day, since January 13. These trends show a considerable drop in requests in most countries as the virus started spreading and governments took action, enforcing or requiring social distancing. Similarly, the number of requests is gradually increasing since June/July.
For flights we use the OpenSky COVID-19 Flight Dataset \footnote{\href{https://opensky-network.org/community/blog/item/6-opensky-covid-19-flight-dataset}{https://opensky-network.org/community/blog/item/6-opensky-covid-19-flight-dataset}}. The OpenSky Network is a non-profit association which provides open access to real-world air traffic control data for research \cite{schafer2014bringing}. Due to high demand, they created a data set with detailed information about all flights in 2019 and 2020. It was made publicly available in April 2020 and is currently updated monthly. The data set is amongst others being used by the Bank of England to now-cast economic indicators during the crisis \cite{BankofEngland}. To compare the flights data with the MT data, we count the number of flights to and from each country per day relative to a predefined baseline during the same period, see Section \ref{s:result}. 
In addition to the mobility data, we look at the number of confirmed cases and deaths per day in each country. These data originate from the World Health Organization\footnote{See: \href{https://covid19.who.int/table}{https://covid19.who.int/table}}.
Finally, for each of the considered countries, we gathered information about the policies taken, such as when a lock-down was put in place. These data came from country specific official information websites.

The time frame we look at in the data starts on January 13, which is the first day available in the MT data set, and the data presented herein ends on July 31, as the flight data is released in monthly batches. Overall we have over six months of data reported on a daily basis.

For the purpose of this paper, we chose to consider only countries in Europe in addition to New Zealand, where response to the pandemic was fast and policies were very effective.
With this data we created a mobility dashboard using the \texttt{shiny} package in R. The mobility dashboard will be available online upon publication of this paper\footnote{here we put a link}. 

\section*{Results}\label{s:result}
We are interested in knowing more about general mobility trends and well-being in Europe during the pandemic. The following analysis is based on data from Apple (as described above) as well as from the flight data set (also described above). The data is compiled in our dashboard.

\subsection*{Mobility Trends}
\begin{figure}
    \centering
    \includegraphics[scale=0.4]{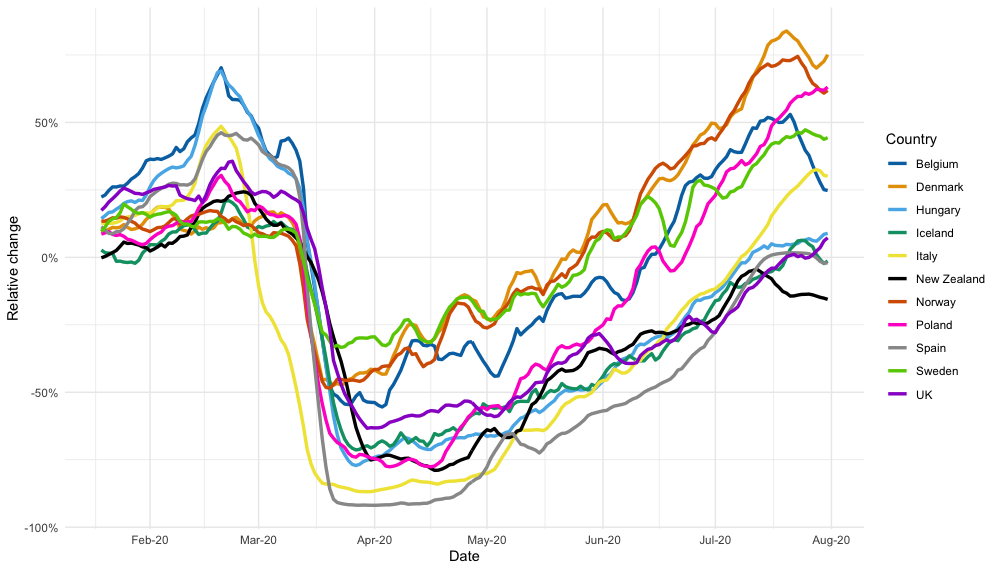}
    \caption{Changes is walking mobility of selected countries}
    \label{fig:walking}
\end{figure}

\begin{figure}
    \centering
    \includegraphics[scale=0.4]{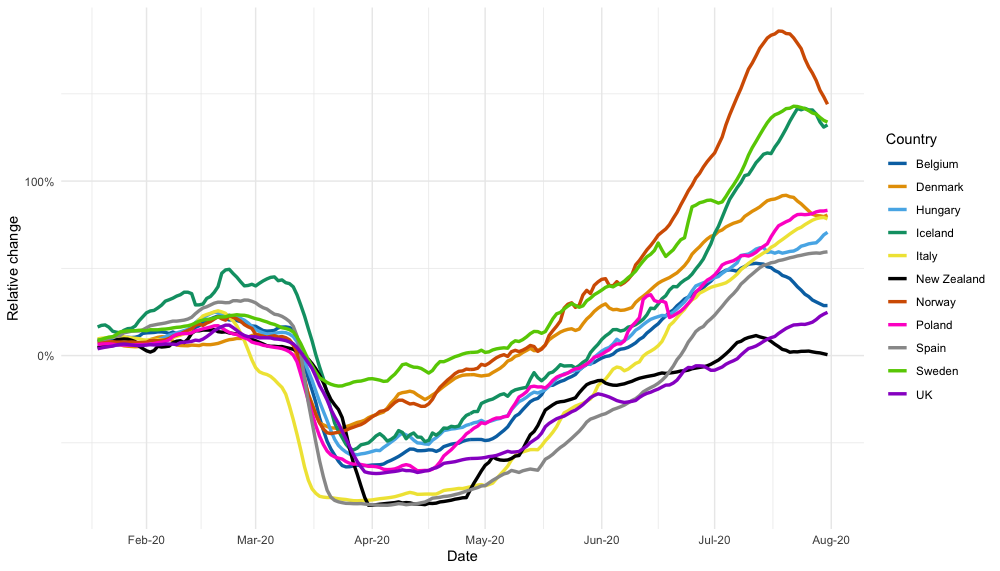}
    \caption{Changes is driving mobility of selected countries}
    \label{fig:driving}
\end{figure}

\begin{figure}
    \centering
    \includegraphics[scale=0.4]{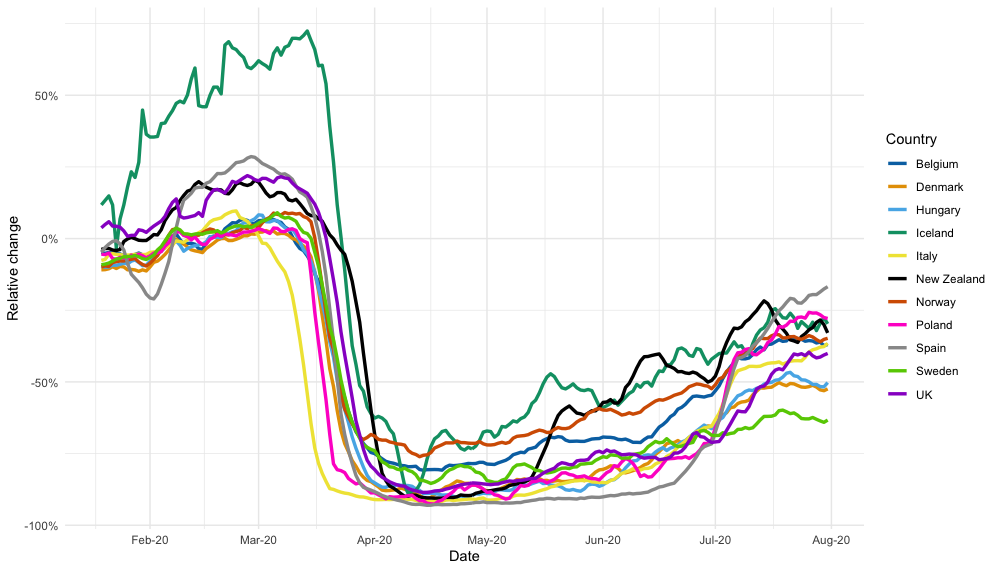}
    \caption{Changes is flying mobility of selected countries}
    \label{fig:flying}
\end{figure}

Figs. \ref{fig:walking}, \ref{fig:driving} and \ref{fig:flying} show, respectively, the changes in walking, driving and flying mobility in ten European countries as well as in New Zealand (black line). The figures show a seven-day moving average of the daily mobility given in the MT and flight datasets. The baseline (0\%) is the mobility of the respective country on the first day of the period.

A general conclusion that can be drawn from this data is that there is a significant decrease in all types of mobility in Europe in March 2020, which gradually increases again although some countries have to date, not gotten back to their normal level of physical activity. 
Flights are still under the baseline, whereas driving has instead increased significantly. 

Notably, the mobility in Italy (yellow line) dropped a few days before the other countries.
The countries Italy and Spain (grey) drop lower in walking and driving mobility than the other countries, which can be explained by the strict lock-down which was imposed in these countries due to the rapid spread of the virus in March. Clearly, their mobility is also recovering more slowly, since restrictions were only lifted gradually.
Sweden (light green) shows a different behaviour from the other countries. Their walking and driving mobility dropped less and recovered faster. 
The other Scandinavian countries, Denmark (light orange) and Norway (dark orange) have rather similar curves to begin with since they responded to the pandemic with similar measures at a similar time, but we see that mobility in Norway is recovering faster than in Denmark, especially regarding driving and flying.

At the end of February, there is small peak in the flying mobility, which reflects the number of people that were rushing to reach their destinations or home countries before restrictions were imposed at a global scale.

Finally, although the pandemic started residing, at least during the summer months in Europe, the imminent second wave and reaction to it, shows that mobility is trending down at the end of July, at least in some of the countries.

\subsection*{Action \& Reaction}
We now attempt to put the mobility trends in perspective with the threat imposed by the spread of the virus and the actions taken by the various governments to contain it.  
Table \ref{T:days} shows important dates for each of the eleven countries we focus on in this paper. These are the date of the first confirmed infection, the date on which a lock-down was imposed, the date where the number of confirmed cases started decreasing and the date of the first death caused by the virus. In addition, the number of days from the first confirmed infection, is shown in parenthesis. This roughly shows how quickly governments responded, how long it took to slow down the spread of the virus and how quickly it escalated.

\begin{table}[!ht]
\begin{adjustwidth}{-2.25in}{0in} 
\centering
\caption{\bf{Important dates by country. The numbers in parenthesis () show the number of days from the first confirmed case in each country.}}
\begin{tabular}{p{2.5cm}p{2cm}p{2.7cm}p{3cm}p{3cm}} \hline
Country&Date of first confirmed case& Date of lock-down (\#Days)& Date of downwards trend (\#Days)& Date of first death (\#Days) \\\hline
 Belgium&February 4&March 18 (43)&April 12 (68)&March 11 (36)\\
 Denmark&February 27&March 13 (15)&April 9 (42)&March 14 (16)\\
 Hungary&March 5&March 30 (25)&April 11 (37)&March 20 (15)\\
 Iceland&February 28&March 16 (17)&April 4 (36)&March 21 (22)\\
 Italy&January 31&March 9 (38)&March 28 (57)&February 24 (24)\\
 Norway&February 26&March 12 (15)&March 29 (32)&March 12 (15)\\
 Poland&March 4&March 13 (9)&--\textsuperscript{b}&March 13 (9)\\
 Spain&January 31&March 14 (43)&April 4 (64)&March 6 (35)\\
 Sweden&January 31&--\textsuperscript{a}&June 24 (145)&March 11 (40)\\
 United Kingdom&January 31&March 23 (52)&May 8 (98)&March 10 (39)\\
 New Zealand & February 28&March 23 (24)&March 29 (30)&March 28 (29) \\ \hline
\end{tabular}
\begin{flushleft} 
(a) Sweden never imposed any form of a lock-down.\\
(b) The number of confirmed cases is still rising in Poland.
\end{flushleft}
\label{T:days}
\end{adjustwidth}
\end{table}

In addition, Table \ref{T:cases} shows the amount of cases at the end of the time period we consider in this paper.  The table shows the number of confirmed infections, the number per one hundred thousand inhabitants, the confirmed number of deaths and the confirmed number of deaths per one hundred thousand inhabitants.

\begin{table}[!ht]
\begin{adjustwidth}{-2.25in}{0in} 
\centering
\caption{\bf{Number of confirmed cases and deaths}}
\begin{tabular}{lp{4cm}p{4cm}} \hline
Country& Total number of infections (per 100k)&Total number of deaths (per 100k) \\\hline
 Belgium&69756 (609)&9845 (86) \\
 Denmark&13789 (237)&615 (11)\\
 Hungary&4535 (46)&597 (6)\\
 Iceland&1907 (534)&10 (3)\\
 Italy&248070 (411)&35154 (58)\\
 Norway&9208 (173)&255 (5)\\
 Poland&46894 (123)&1731 (5)\\
 Spain&288522 (615)&28445 (61)\\
 Sweden&80422 (786)&5743 (56)\\
 United Kingdom&304699 (457)&46201 (69)\\
 New Zealand&1217 (25)&22 (0.5) \\ \hline
\end{tabular}
\label{T:cases}
\end{adjustwidth}
\end{table}

Comparing countries that impose a lock-down before and after they experience the first death due to the virus, we have on one hand Denmark, Iceland, Norway, Poland and New Zealand, and on the other hand, Belgium, Hungary, Italy, Spain, Sweden and the United Kingdom.
In the first group, there have been fewer deaths per 100k inhabitants as well as fewer number of confirmed infections per 100k inhabitants (excluding Hungary and Iceland).
For the countries in the first group, less than 50 days passed before the number of daily new infections started to decrease, where as the countries that took longer to react also had to wait longer for the virus to start residing.

It is worth mentioning that we are only looking at lock-downs as a form of government response. However, alongside this, governments insisted on social distancing, hand washing, ban on social gatherings over a certain size, closing of schools etc. that also had an effect on the spread. 
What is clear from these observations, however, is that timely response was crucial.
This is inline with \cite{hsiang2020effect}.

\section*{Discussion}\label{s:discussion}
Time plays a pivotal role in a crisis situation, and that is also true about this particular pandemic \cite{aagerfalk2020information}. Using open source data, and pulling together data from different sources which shows the importance of when actions were taken and to what extent, and we see that as an important contribution in the wake of the pandemic. Technology and the development of information systems such as the dashboard presented herein which relies on open source data and shows actions in a transparent way, is, as we see it, at the essence of our role as researchers, alongside the ability to analyse vast amounts of data around the unprecedented upheaval that COVID-19 has posed. Those two aspects thereby outline the main contribution of this paper. 
There are several insights we can obtain from the data presented in the previous section and here we will discuss the following aspects: i) changes in mobility trends, ii) flying versus driving, iii) specific countries in relation to the level of harshness in policy taken across Europe iv) long term effects on well-being and the role of habit and rules on everyday activity, v) the use of open source data and the importance of granularity and accuracy of open data.

A general conclusion that can be drawn from our analyses of the open source data is that there is a significant decrease in all types of mobility in Europe. The drop starts in March 2020, and gradually increases again although some countries have to date, not gotten back to their normal level of physical activity. Regarding flights, they are to date still under the baseline. The sudden drop in mobility is not surprising given the heavy restrictions in most countries. What is more interesting is how differently these trends increase again.  

Even though flying has decreased significantly, driving has in turn increased, especially after the ease on travel restrictions. An assumption that can be made is thereby that even though we are flying less, we are not really travelling less, or at least we are not catering to the sustainable aspect of flying less in general. Instead we compensate flying less with driving more. Behavioural changes like travelling less, and acting in an increasingly sustainable way in everyday activities are rooted in our human behaviour and a pandemic can staple that type of behaviour short term. In contrast, acting in an increasingly sustainable way long term, is something that needs to happen in smaller steps over a longer period of time. Travel restrictions were eased in Europe with summer and summer holidays around the corner, so it is logical to assume that people drove to their holiday destination instead of flying. To shed further light on this, it would be interesting to know where people went on holiday, that is, how far away from their homes.

Notably,  the  mobility  in  Italy dropped  a  few  days  before the  other countries. The  countries  Italy  and  Spain drop lower in  walking and  riving mobility than the other countries over the period examined, which can be explained by the strict lock-down which was imposed in these countries due to the rapid spread of the virus in March. Clearly, their mobility is also recovering slower, since restrictions were only lifted slowly. In the data, Sweden shows a different behaviour in comparison to the other countries. Sweden did not use lock-down strategies and as a result, the data shows a different trend. Their walking and driving mobility dropped less and recovered faster. The other Scandinavian countries show rather similar curves to begin with, as their respond to the pandemic and their strategy included similar measures at a similar time whereas the mobility in Norway recovering faster than in Denmark, especially when looking towards driving and flying. This particular aspect --which strategy is the best-- will likely remain an open research question for a long time after the pandemic has passed. Even though lock-downs and social distancing are accepted as effective measures, we can not know what the long term economic and public health consequences will be \cite{hsiang2020effect}. 

As mentioned above, the long term effects of the pandemic remain unknown.  Millions of people were restricted to their homes for weeks, which caused drastic changes in daily routines as we have demonstrated with the mobility data. In addition, according to open source data from Withings\footnote{Withings produces activity tracking devices such as watches and Bluetooth scales and compiled weight data from the pandemic time: \href{https://blog.withings.com/tag/health-data/}{https://blog.withings.com/tag/health-data/}}, the average person in France has gained .19 lbs (.084 kg), in Spain .26 lbs (.117 kg), in United Kingdom .35 lbs (.16 kg), in Germany .41 lbs (.189 kg), in Italy .42 lbs (.195 kg) and in China .55 lbs (.25 kg). Data on steps taken on a daily basis from China shows a decrease by over half (56\%) in Hubei during the lock-down period. According to the same data set from Withings, daily step count decreased by 8\% in the United Kingdom, by 28\% in Italy, 27\% in France and an increase of 1\% during self-isolation in Germany. What this data shows, is that the impact of the measures taken, differs greatly in the countries listed here above and bear in mind that this data is only from Withings.
This data from Withings, in combination with the mobility data shows that during the lock-down period, there is a short-term effect on weight, and a connection between reduced walking. The data from Withings is however only from a short period of time and in additon to that, the numbers are based on data from a limited part of the population (only those that own a Withings wearable) whereas the long-term effects on well-being in general need to be studied further, using larger data sets over an extended period of time.

The analysis herein, which is based on open and publicly available data, shows the importance of opening up the data that is already being gathered across different platforms. What we would therefore like to emphasise is the importance of granularity and accuracy of open source data. 
There is much added value in combining data from different sources, and presenting it in an intuitive and clear manner, as we have done here. Links and correlations that were unknown before can easily be seen and interpreted. 
As data privacy and ethics of data usage are currently widely being discussed and disputed, it is important to emphasise that data in its most aggregate form --as we used here-- can come to great use, when presented correctly.
With the vast and widespread data collection that is taking place at every front, for example by social media platforms, there is a call for stronger incentives for these organizations to share insights and preserve historically significant data \cite{ohman2020what}.

\section*{Conclusion}\label{s:conclusion}
The COVID-19 pandemic has triggered a worldwide health crisis and the strategies taken across Europe have differed where some countries have taken strict measures, while others have avoided lock-downs. In this paper we report on findings from putting together open data from different sources in order to shed light on the changes in mobility patterns in Europe during the pandemic. Using that data we show the impact of the different strategies put into place in Europe during the pandemic and how mobility patterns changed in different counties. Our data shows how the majority of Europe walked less during the lock-downs, which is bound to have a long-term effect on well-being. Also, while flights were less frequent, driving increased drastically. Through our analysis we show the importance of data granularity and accuracy through our novel study of using open source data to shed light on mobility trends in the pandemic of COVID-19.

Our research offers several opportunities for future work.  As the summary statistics about weight gain during lock-down indicate, analysing other sources of data alongside what we do in this paper would be an interesting way of continuing this line of work. Especially in regards to data describing health and well-being in order to examine the long-term effects of the pandemic.
The OpenSky data set only provided information about flights, not the quantity of people on each flight. These numbers could be interesting to take into account, since during the travel restrictions, relatively few people were on each flight. 
In our results, we described the variation in the impact of the pandemics with respect to temporal factors, such as the delay between first infection and imposing of lock-downs in Europe. In our future work, we will investigate such correlations to a greater extent at a global level in order to contribute to an understanding what measures were the most effective when designing strategies during a fight against a pandemic.

\nolinenumbers

%
%
%


\begin{thebibliography}{10}
\bibitem{wu2020new}
Wu F, Zhao S, Yu B, Chen YM, Wang W, Song ZG, et~al.
\newblock A new coronavirus associated with human respiratory disease in China.
\newblock Nature. 2020;579(7798):265--269.

\bibitem{huang2020clinical}
Huang C, Wang Y, Li X, Ren L, Zhao J, Hu Y, et~al.
\newblock Clinical features of patients infected with 2019 novel coronavirus in
  Wuhan, China.
\newblock The lancet. 2020;395(10223):497--506.

\bibitem{zhu2020novel}
Zhu N, Zhang D, Wang W, Li X, Yang B, Song J, et~al.
\newblock A novel coronavirus from patients with pneumonia in China, 2019.
\newblock New England Journal of Medicine. 2020;.

\bibitem{shangguan2020caused}
Shangguan Z, Wang MY, Sun W.
\newblock What caused the outbreak of COVID-19 in China: From the perspective
  of crisis management.
\newblock International Journal of Environmental Research and Public Health.
  2020;17(9):3279.

\bibitem{kissler2020social}
Kissler SM, Tedijanto C, Lipsitch M, Grad Y.
\newblock Social distancing strategies for curbing the COVID-19 epidemic.
\newblock medRxiv. 2020;.

\bibitem{hsiang2020effect}
Hsiang S, Allen D, Annan-Phan S, Bell K, Bolliger I, Chong T, et~al.
\newblock The effect of large-scale anti-contagion policies on the COVID-19
  pandemic.
\newblock Nature. 2020; p. 1--9.

\bibitem{braa2004networks}
Braa J, Monteiro E, Sahay S.
\newblock Networks of action: sustainable health information systems across
  developing countries.
\newblock MIS quarterly. 2004; p. 337--362.

\bibitem{aagerfalk2020information}
{\AA}gerfalk PJ, Conboy K, Myers MD. Information systems in the age of
  pandemics: COVID-19 and beyond; 2020.

\bibitem{heeks2010social}
Heeks R, Arun S.
\newblock Social outsourcing as a development tool: The impact of outsourcing
  IT services to women's social enterprises in Kerala.
\newblock Journal of International Development: The Journal of the Development
  Studies Association. 2010;22(4):441--454.

\bibitem{rajao2009conceptions}
Raj{\~a}o R, Hayes N.
\newblock Conceptions of control and IT artefacts: an institutional account of
  the Amazon rainforest monitoring system.
\newblock Journal of Information Technology. 2009;24(4):320--331.

\bibitem{tim2017digitally}
Tim Y, Pan SL, Ractham P, Kaewkitipong L.
\newblock Digitally enabled disaster response: the emergence of social media as
  boundary objects in a flooding disaster.
\newblock Information Systems Journal. 2017;27(2):197--232.

\bibitem{islind2018co}
Islind AS, Lundh~Snis U.
\newblock From co-design to co-care: designing a collaborative practice in
  care.
\newblock Systems, Signs \& Actions. 2018;11(1):1--24.

\bibitem{pan2020fighting}
Pan SL, Zhang S.
\newblock From fighting COVID-19 pandemic to tackling sustainable development
  goals: An opportunity for responsible information systems research.
\newblock International Journal of Information Management. 2020; p. 102196.

\bibitem{buckee2020aggregated}
Buckee CO, Balsari S, Chan J, Crosas M, Dominici F, Gasser U, et~al.
\newblock Aggregated mobility data could help fight COVID-19.
\newblock Science (New York, NY). 2020;368(6487):145.

\bibitem{gonzalez2008understanding}
Gonzalez MC, Hidalgo CA, Barabasi AL.
\newblock Understanding individual human mobility patterns.
\newblock nature. 2008;453(7196):779--782.

\bibitem{oskarsdottir2020interaction}
{\'O}skarsd{\'o}ttir M, {\'I}slind AS.
\newblock Interaction with Medical Venues in Megacities.
\newblock Studies in health technology and informatics. 2020;270:996--1000.

\bibitem{wilson2016rapid}
Wilson R, zu~Erbach-Schoenberg E, Albert M, Power D, Tudge S, Gonzalez M,
  et~al.
\newblock Rapid and near real-time assessments of population displacement using
  mobile phone data following disasters: the 2015 Nepal Earthquake.
\newblock PLoS currents. 2016;8.

\bibitem{balcan2010modeling}
Balcan D, Gon{\c{c}}alves B, Hu H, Ramasco JJ, Colizza V, Vespignani A.
\newblock Modeling the spatial spread of infectious diseases: The GLobal
  Epidemic and Mobility computational model.
\newblock Journal of computational science. 2010;1(3):132--145.

\bibitem{bajardi2011human}
Bajardi P, Poletto C, Ramasco JJ, Tizzoni M, Colizza V, Vespignani A.
\newblock Human mobility networks, travel restrictions, and the global spread
  of 2009 H1N1 pandemic.
\newblock PloS one. 2011;6(1):e16591.

\bibitem{kishore2020flying}
Kishore N, Mitchell R, Lash TL, Reed C, Danon L, Sigmundsd{\'o}ttir G, et~al.
\newblock Flying, phones and flu: Anonymized call records suggest that Keflavik
  International Airport introduced pandemic H1N1 into Iceland in 2009.
\newblock Influenza and other respiratory viruses. 2020;14(1):37--45.

\bibitem{tang2020estimation}
Tang B, Wang X, Li Q, Bragazzi NL, Tang S, Xiao Y, et~al.
\newblock Estimation of the transmission risk of the 2019-nCoV and its
  implication for public health interventions.
\newblock Journal of clinical medicine. 2020;9(2):462.

\bibitem{cao2020incorporating}
Cao Z, Zhang Q, Lu X, Pfeiffer D, Wang L, Song H, et~al.
\newblock Incorporating human movement data to improve epidemiological
  estimates for 2019-nCoV.
\newblock medRxiv. 2020;.

\bibitem{tian2020investigation}
Tian H, Liu Y, Li Y, Wu CH, Chen B, Kraemer MU, et~al.
\newblock An investigation of transmission control measures during the first 50
  days of the COVID-19 epidemic in China.
\newblock Science. 2020;368(6491):638--642.

\bibitem{kraemer2020effect}
Kraemer MU, Yang CH, Gutierrez B, Wu CH, Klein B, Pigott DM, et~al.
\newblock The effect of human mobility and control measures on the COVID-19
  epidemic in China.
\newblock Science. 2020;368(6490):493--497.

\bibitem{chinazzi2020effect}
Chinazzi M, Davis JT, Ajelli M, Gioannini C, Litvinova M, Merler S, et~al.
\newblock The effect of travel restrictions on the spread of the 2019 novel
  coronavirus (COVID-19) outbreak.
\newblock Science. 2020;368(6489):395--400.

\bibitem{aleta2020modelling}
Aleta A, Mart{\'\i}n-Corral D, y~Piontti AP, Ajelli M, Litvinova M, Chinazzi M,
  et~al.
\newblock Modelling the impact of testing, contact tracing and household
  quarantine on second waves of COVID-19.
\newblock Nature Human Behaviour. 2020; p. 1--8.

\bibitem{warren2020mobility}
Warren MS, Skillman SW.
\newblock Mobility changes in response to COVID-19.
\newblock arXiv preprint arXiv:200314228. 2020;.

\bibitem{carteni2020mobility}
Carten{\`\i} A, Di~Francesco L, Martino M.
\newblock How mobility habits influenced the spread of the COVID-19 pandemic:
  Results from the Italian case study.
\newblock Science of The Total Environment. 2020; p. 140489.

\bibitem{jia2020population}
Jia JS, Lu X, Yuan Y, Xu G, Jia J, Christakis NA.
\newblock Population flow drives spatio-temporal distribution of COVID-19 in
  China.
\newblock Nature. 2020; p. 1--5.

\bibitem{schmidt2013behaviour}
Schmidt S, Galea E.
\newblock Behaviour-security-culture human behaviour in emergencies and
  disasters: a cross-cultural investigation.
\newblock Pabst Science Publishers; 2013.

\bibitem{karanasios2020gatekeepers}
Karanasios S, Cooper V, Adrot A, Mercieca B.
\newblock Gatekeepers Rather than Helpless: An Exploratory Investigation of
  Seniors’ Use of Information and Communication Technology in Critical
  Settings.
\newblock In: Proceedings of the 53rd Hawaii International Conference on System
  Sciences; 2020.

\bibitem{spinelli2020covid}
Spinelli A, Pellino G.
\newblock COVID-19 pandemic: perspectives on an unfolding crisis.
\newblock The British journal of surgery. 2020;.

\bibitem{flaxman2020estimating}
Flaxman S, Mishra S, Gandy A, Unwin HJT, Mellan TA, Coupland H, et~al.
\newblock Estimating the effects of non-pharmaceutical interventions on
  COVID-19 in Europe.
\newblock Nature. 2020; p. 1--5.

\bibitem{schafer2014bringing}
Sch{\"a}fer M, Strohmeier M, Lenders V, Martinovic I, Wilhelm M.
\newblock Bringing up OpenSky: A large-scale ADS-B sensor network for research.
\newblock In: IPSN-14 Proceedings of the 13th International Symposium on
  Information Processing in Sensor Networks. IEEE; 2014. p. 83--94.

\bibitem{BankofEngland}
{Bank of England}. Monetary Policy Report and Interim Financial Stability
  Report - May 2020; 2020.
\newblock
  https://www.bankofengland.co.uk/report/2020/monetary-policy-report-financial-stability-report-may-2020.

\bibitem{ohman2020what}
Öhman C, Aggarwa N.
\newblock What if Facebook goes down? Ethical and legal considerations for the
  demise of big tech.
\newblock Internet Policy Review,. 2020;9(3).
\newblock doi:{10.14763/2020.3.1488}.


\end{thebibliography}
\end{document}